# Chaotic dynamics of movements stochastic instability and the hypothesis of N.A. Bernstein about "repetition without repetition"

Eskov V.V.[1], Volov V.T.[2], Eskov V.M.[1], Ilyashenko L.K.[3]

[1] Scientific research institute of system analysis

[2] Department of Natural sciences, Samara State Transport University, Samara, Russia

[3] Department of Natural and Humanities Sciences, Tyumen Industrial University,

**Abstract.** The registration of tremor was performed in two groups of subjects (15 people in each group) with different physical fitness at rest and at a static loads of 3*N*. Each subject has been tested 15 series (number of series *N*=15) in both states (with and without physical loads) and each series contained 15 samples (*n*=15) of tremorogramm measurements (500 elements in each sample, registered coordinates $x_1(t)$ of the finger position relative to eddy current sensor) of the finger. Using non-parametric Wilcoxon test of each series of experiment a pairwise comparison was made forming 15 tables in which the results of calculation of pairwise comparison was presented as a matrix (15x15) for tremorogramms are presented. The average number of hits random pairs of samples ($<k>$) and standard deviation σ were calculated for all 15 matrices without load and under the impact of physical load (3N), which showed an increase almost in twice in the number *k* of pairs of matching samples of tremorogramms at conditions of a static load. For all these samples it was calculated special quasi-attractor (this square was presented the distinguishes between physical load and without it. All samples present the stochastic unstable state

## INTRODUCTION

The human brain and the motor system solve the challenging task of building a movement for which it is not enough simply counting the number of mechanical degrees of freedom (number of joints and muscles in the hand). Most of the actions of neuromuscular systems (NMS) are dynamic and require continuous and coordinated work of all elements of the system, which in the end, we will show, working chaotically [1-7]. The chaos of samples (under numerical repetition) presents the uninterrupted changes of its statistical distribution function $f(x)$ (so $f_j(x_i) \neq f_{j+1}(x_i)$ for any $j$-th and $j+1$-th samples) [8-15].

In our studies the parameters of a NMS of the person have been analyzed, such parameters which characterizes the change of the parameters of the chaotic NMS (for women and man) when performing regular exercise (when compared with the rest of the subjects, without physical training).

The traditional understanding of stationary regimes of biological systems in the deterministic form $dx/dt=0$, where $x=x(t)=(x_1,x_2,...,x_n)^T$ is a vector of system state, or when calculating the distribution functions $f(x)$ when the stationary regime requires the stability of these $f(x)$, obtained for consecutive samples of a parameter $x$ is not useful. We use the pairwise comparisons of such samples and construct the matrix of samples [11-15]. Real human motions are chaotic, i.e. $dx/dt \neq 0$ constantly, i.e. it is almost impossible to obtain two adjacent samples $f_j(x_i(t))=f_{j+1}(x_i(t))$. In this regard, proposed and new methods for the calculation of chaotic dynamics of tremor (as alleged involuntary movements) and not only tremor [1, 9, 16-21].

The aim of this study is to assess the peculiarities of chaotic dynamics of tremor micro-movements of the human upper limb with different physical preparedness (without load and under conditions of static loads) from the position of new theory of chaos-self-organization – TCS and Eskov-Zinchenko effect [1-8, 11-16].

**Materials and Methods.** Parameters of tremor of the subjects were recorded using biophysical measuring complex, developed in the laboratory of bio-cybernetics and biophysics of complex systems in Surgut state university (fig.1). The installation includes a metal plate (2) which is fixed rigidly to the finger test, eddy-current sensor (1), the amplifier together with an analog-to-digital converter (ADC) (3) and the computer with the original software (4) [17-20].

As a first phase coordinate in addition to coordinate $x_1=x_i(t)$ of the moving limbs we used the second coordinate as a speed of movement of the finger $x_2=dx_1/dt$ [4, 6]. Before the subjects had the task to hold the finger in a given area, consciously controlling his immobility at a given point of the space. We investigated two group of testing object: 12 woman (with average age $<T_1>$ = 29 years) and 12 man ($<T_2>$=27 year) according to Helsinki Declaration.

Each subject took 15 episodes (series) of the experiments ($N$=15), in each of which the registration of tremor was performed 15 times ($n$=15) at rest and similarly ($N$=15, n=15) under a load of 3$N$ (load, attached to the index finger).

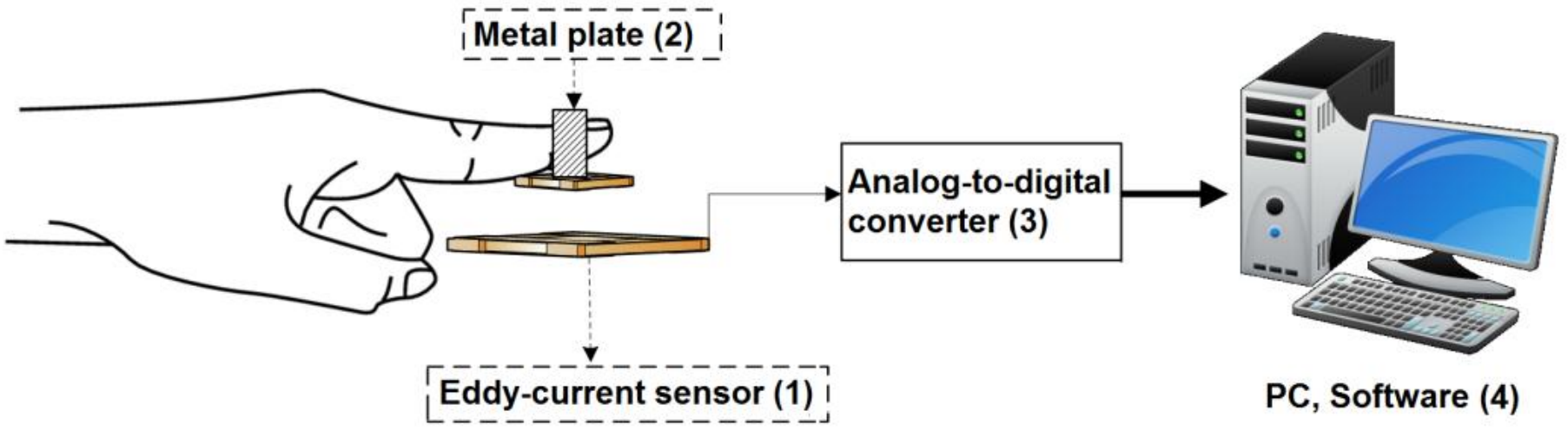

Fig.1. Scheme of measuring complex for tremor to be recorded

**Statistical and chaotic analysis.** Statistical data processing was carried out using the software package "Statistica 10". Analysis of compliance distributions of received data to a normal distribution was based on calculation of the Shapiro-Wilk test. A Wilcoxon test has been used to perform non-parametric pairwise comparison of tremorogramms (TMG) and 15 tables for each test subject in a resting state and 15 in the conditions of a static load of 3N for each test subject (a total of 225 samples of TMG) has been built.

Using ADC (3) and TMG have been recorded in the file with sampling rate $t=0.01$ seconds (total time of registration of the $i$-th sample of $T$=5 sec., the number of points in the expanded file $z$=500). It was then produced pairwise comparison of segments of TMG for TMG sample of each test subject in order to classify all these samples as samples of the one general population (for the same test subject under a certain homeostatic state) [28].

Data processing and recording of tremor of the extremities of the subjects were carried out with the help of a computer (the program «Charts 3» was used»). Due to the patented software product, the phase planes were constructed and the areas S were calculated in the form $S = \Delta x_1 * \Delta x_2$, where $\Delta x_1$ is the variation range for TMG by $x_1$, $\Delta x_2$ is the variation range by $x_2$. In addition, a pairwise comparison of the segments of the tremorograms was performed for each sample $x_1 = x_1 (t)$ of the subject's tremorogramms to determine whether all of these samples belonged to the general population (for same subject). The comparison was made by calculating the matrices of pair comparisons of samples $x_1$.

. Conformity analysis of type of distribution of received data for normal distribution law was performed on the basis of the Shapiro-Wilk test calculation. It was found out that the Gaussian distribution can not be applied to all samples of TMG. Then, using a nonparametric pair comparison of tremorogramms with the Wilcoxon test, 30 tables of paired comparisons of TMG samples were constructed for each subject in the form of 15 series of experiments. Each series had 15 comparative samples of TMG in a calm state and after local cold exposure. Simultaneously parameters of QA (in particular areas) were calculated for the same samples of TMG.

## RESULTS OF THE STUDY

A particular chaotic dynamics of involuntary micro-movements of the limbs (tremor of the fingers), as a reaction to the static load, manifested in the change in the number of matches arbitrary pairs of samples ($k$), (pair) include one of the General population. For this case the matrix of pairwise comparisons has been calculated ($N$=15 without load, $F_1$=0 and with load, $F_2=3N$ for every woman and every man (all group $N_1$ consist of 12 women and $N_2$ consist of 12 men)). Note that Eskov-Zinchenko effect declares the lack of statistical robustness (there are no coincidences in a row of the received samples $x_i$) for any of homeostasic parameters of biological systems. We deal with NMS, for which we identify tremor as an involuntary movement. Let us demonstrate this claim in the mode of multiple repetitions of registration of tremor (as a sample) in the form of 2 matrices of all 60 obtained matrices (as a model) for the female and 2 matrices for the male (under 225 repetition without load and 225 repetitions without load and 225 repetitions with load $F_2$=3$N$ as for the woman and as for the man) [9, 11-13].

Paired comparisons of samples of TMG obtained in each series ($N$=1) of $n$=15 iterations of TMG registration (500 points – the values of $x_i(t)$ coordinate of the finger in relation to the eddy current sensor check tremor) are presented in tables 1-4. Specific ($N$=1) examples of calculation results of matrices (15x15) for pairwise comparison of TMG (coordinate $x_i(t)$) of subjects with different physical training showed that the number of pairs of identical samples is small ($k_{11}$=3, $k_{12}$=6, $k_{21}$=3, $k_{22}$=9, but they differ greatly as an athlete and person without physical training.

As an example, Table 1 presents the typical matrix for women as a test subject – FDU (according to 225 TMG samples, using this method 15 matrices has been built for each test subject) in case of the resting state (free stat, $F_1$=0). At the same time, for each such resting state, the experiment was repeated, but with physical load of $F_2$=3$N$, 225 TMG samples of the same test subject has been recorded. Physical load on the finger (static load, 3$N$) allowed us to calculate 15 matrices (with physical load). A typical example of this (the second parallel) experiment has been shown in table 2.

A pairwise comparisons of TMG samples of man as a test subject – GDV has been made. A typical example of pairwise matrices calculation was presented in Tables 3 and 4. A comparison of the data for this test subject showed an increase in the number of matches $k$ pairs of samples 2 times at a physical load (3$N$): $k_{21}$=3 (no physical load) and $k_{22}$=9 (with load of 3$N$). In this case there is no element in top-diagonal set $p$, where $p$<0.01. It means a sharp decline in the share of stochastics in TMG evaluation (there is no possibility to register two consecutive TMG samples with $p$<0.01) Adjacent pairs of TMG are stochastically unique in case of athletes $f_j(x_i)\neq f_{j+1}(x_i)$.

*Table 1*

**The matrix of pairwise comparison of the woman (FDU) test tremorogramm (no load, number of repetitions $n$=15), was used the Wilcoxon test (significance $p$<0.05, the number of matches $k_{11}$=3)**

| | 1 | 2 | 3 | 4 | 5 | 6 | 7 | 8 | 9 | 10 | 11 | 12 | 13 | 14 | 15 |
|---|---|---|---|---|---|---|---|---|---|---|---|---|---|---|---|
| 1 | | .00 | .00 | .00 | .00 | .00 | .00 | .00 | **.63** | .00 | .00 | .00 | .00 | .00 | .00 |
| 2 | .00 | | .00 | .00 | .00 | .00 | .00 | .00 | .00 | .00 | .00 | .00 | .00 | .00 | .00 |
| 3 | .00 | .00 | | .00 | .00 | .00 | **.69** | .00 | .00 | .00 | .00 | .00 | .00 | .00 | .00 |
| 4 | .00 | .00 | .00 | | .00 | .00 | .00 | .00 | .00 | .00 | .00 | .00 | .00 | .00 | .00 |
| 5 | .00 | .00 | .00 | .00 | | .00 | .00 | .00 | .00 | .00 | .00 | .00 | .00 | .00 | .00 |
| 6 | .00 | .00 | .00 | .00 | .00 | | .00 | .00 | .00 | .00 | .00 | .00 | .00 | .00 | .00 |
| 7 | .00 | .00 | **.69** | .00 | .00 | .00 | | .00 | .00 | .00 | .00 | .00 | .00 | .00 | .00 |
| 8 | .00 | .00 | .00 | .00 | .00 | .00 | .00 | | .00 | .00 | .00 | .00 | .00 | .00 | .00 |
| 9 | **.63** | .00 | .00 | .00 | .00 | .00 | .00 | .00 | | .00 | .00 | .00 | .00 | .00 | .00 |
| 10 | .00 | .00 | .00 | .00 | .00 | .00 | .00 | .00 | .00 | | .00 | .00 | .00 | .00 | .00 |
| 11 | .00 | .00 | .00 | .00 | .00 | .00 | .00 | .00 | .00 | .00 | | .00 | .00 | .00 | **.70** |
| 12 | .00 | .00 | .00 | .00 | .00 | .00 | .00 | .00 | .00 | .00 | .00 | | .00 | .00 | .00 |
| 13 | .00 | .00 | .00 | .00 | .00 | .00 | .00 | .00 | .00 | .00 | .00 | .00 | | .00 | .00 |
| 14 | .00 | .00 | .00 | .00 | .00 | .00 | .00 | .00 | .00 | .00 | .00 | .00 | .00 | | .00 |
| 15 | .00 | .00 | .00 | .00 | .00 | .00 | .00 | .00 | .00 | .00 | **.70** | .00 | .00 | .00 | |

*Table 2*

**The matrix of pairwise comparison of the woman (FDU) test tremorogramm (with physical load $F$=3$N$), the number of replications $n$=15), the Wilcoxon test was used (significance level $p$<0.05, number of hits $k_{12}$=6)**

| | 1 | 2 | 3 | 4 | 5 | 6 | 7 | 8 | 9 | 10 | 11 | 12 | 13 | 14 | 15 |
|---|---|---|---|---|---|---|---|---|---|---|---|---|---|---|---|
| 1 | | .00 | .00 | .00 | .00 | .00 | .00 | .00 | .00 | .00 | .00 | .00 | **.47** | .00 | **.24** |
| 2 | .00 | | .00 | .00 | .00 | .00 | .00 | .00 | .00 | .00 | .00 | .00 | .00 | .00 | .00 |
| 3 | .00 | .00 | | **.33** | .00 | .00 | .00 | .00 | .00 | .00 | .00 | .00 | .00 | .00 | .00 |
| 4 | .00 | .00 | **.33** | | .00 | **.71** | .00 | .00 | .00 | .00 | .00 | .00 | .00 | .00 | .00 |
| 5 | .00 | .00 | .00 | .00 | | .00 | .00 | .00 | .00 | .00 | .00 | **.65** | .00 | .00 | .00 |
| 6 | .00 | .00 | .00 | **.71** | .00 | | .00 | .00 | .00 | .00 | .00 | .00 | .00 | .00 | .00 |
| 7 | .00 | .00 | .00 | .00 | .00 | .00 | | .00 | .00 | .00 | .00 | **.52** | .00 | .00 | .00 |
| 8 | .00 | .00 | .00 | .00 | .00 | .00 | .00 | | .00 | .00 | .00 | .00 | .00 | .00 | .00 |
| 9 | .00 | .00 | .00 | .00 | .00 | .00 | .00 | .00 | | .00 | .00 | .00 | .00 | .00 | .00 |
| 10 | .00 | .00 | .00 | .00 | .00 | .00 | .00 | .00 | .00 | | .00 | .00 | .00 | .00 | .00 |
| 11 | .00 | .00 | .00 | .00 | .00 | .00 | .00 | .00 | .00 | .00 | | .00 | .00 | .00 | .00 |
| 12 | .00 | .00 | .00 | .00 | **.65** | .00 | **.52** | .00 | .00 | .00 | .00 | | .00 | .00 | .00 |
| 13 | **.47** | .00 | .00 | .00 | .00 | .00 | .00 | .00 | .00 | .00 | .00 | .00 | | .00 | .02 |
| 14 | .00 | .00 | .00 | .00 | .00 | .00 | .00 | .00 | .00 | .00 | .00 | .00 | .00 | | .00 |
| 15 | **.24** | .00 | .00 | .00 | .00 | .00 | .00 | .00 | .00 | .00 | .00 | .00 | .02 | .00 | |

*Table 3*

**The matrix of pairwise comparison of the man (GDV) test tremorogramm (without physical load, number of repetitions $n$=15), the Wilcoxon test was used (significance level $p$<0.05, the number of matches $k_{21}$=3)**

| | 1 | 2 | 3 | 4 | 5 | 6 | 7 | 8 | 9 | 10 | 11 | 12 | 13 | 14 | 15 |
|---|---|---|---|---|---|---|---|---|---|---|---|---|---|---|---|
| 1 | | .00 | .00 | .00 | .00 | .00 | .00 | .00 | .00 | .00 | .00 | .00 | .00 | .00 | .00 |
| 2 | .00 | | .00 | .00 | .00 | .00 | .00 | .00 | **.45** | .00 | .00 | .00 | .00 | .00 | .00 |
| 3 | .00 | .00 | | .00 | .00 | .00 | .00 | .00 | .00 | **.42** | .00 | .00 | .00 | .00 | .00 |
| 4 | .00 | .00 | .00 | | .00 | .00 | .00 | .00 | .00 | .00 | .00 | .00 | .00 | .00 | .00 |
| 5 | .00 | .00 | .00 | .00 | | .00 | .00 | .00 | .00 | .00 | .00 | .00 | .00 | .00 | .00 |
| 6 | .00 | .00 | .00 | .00 | .00 | | .00 | .00 | .00 | .00 | .00 | .00 | .00 | .00 | .00 |
| 7 | .00 | .00 | .00 | .00 | .00 | .00 | | .00 | .00 | .00 | .00 | .00 | .00 | .00 | .00 |
| 8 | .00 | .00 | .00 | .00 | .00 | .00 | .00 | | .00 | .00 | .00 | .00 | .00 | .00 | .00 |
| 9 | .00 | .45 | .00 | .00 | .00 | .00 | .00 | .00 | | .00 | .00 | .00 | .00 | .00 | .00 |
| 10 | .00 | .00 | .42 | .00 | .00 | .00 | .00 | .00 | .00 | | .00 | .00 | **.17** | .00 | .00 |
| 11 | .00 | .00 | .00 | .00 | .00 | .00 | .00 | .00 | .00 | .00 | | .00 | .00 | .00 | .00 |
| 12 | .00 | .00 | .00 | .00 | .00 | .00 | .00 | .00 | .00 | .00 | .00 | | .00 | .00 | .00 |
| 13 | .00 | .00 | .00 | .00 | .00 | .00 | .00 | .00 | .00 | .17 | .00 | .00 | | .00 | .00 |
| 14 | .00 | .00 | .00 | .00 | .00 | .00 | .00 | .00 | .00 | .00 | .00 | .00 | .00 | | .00 |
| 15 | .00 | .00 | .00 | .00 | .00 | .00 | .00 | .00 | .00 | .00 | .00 | .00 | .00 | .00 | |

The number of matches $k$ pairs of samples in Table 2 ($k_{12}$=6) is greater in 2 times than the number of pairs of comparison samples (for woman) of the same test subject (without physical load, as in Table 1), since $k_{11}$=3. These are examples in case of test subjects without physical training and they show the diagnostic value of calculating the number of $k$ on the background of statistical instability of distribution functions $f(x)$. Indeed, this example demonstrates that for 105 independent pairs of comparisons of tremorogramm samples (see Table 1) only two matched - 3rd and 4th samples ($p$=0.69) in Table 1. In Table 2 we have only one such pair. Moreover, it is impossible to obtain matching pairs of tremorogramms randomly! However, for all 60 matrices the probability of this pair coincidence (in a row) was even less. Everything happens according to N.A. Bernstein principle of "repetition without repetition" [25] and it forms a quantitative estimation of Eskov-Zinchenko effect [1-8].

*Table 4*

**The matrix of pairwise comparison of the man (GDV) test tremorogramm (with physical load ($F$=3$N$), the number of replications $n$=15), the Wilcoxon test was used (significance $p$<0.05, the number of matches $k_{22}$=9)**

| | 1 | 2 | 3 | 4 | 5 | 6 | 7 | 8 | 9 | 10 | 11 | 12 | 13 | 14 | 15 |
|---|---|---|---|---|---|---|---|---|---|---|---|---|---|---|---|
| 1 | | .00 | .00 | .00 | .00 | .00 | .00 | .00 | .00 | .00 | .00 | .00 | .00 | .00 | .00 |
| 2 | .00 | | .00 | .00 | .00 | .00 | .00 | .00 | .00 | **.42** | **.27** | .00 | .00 | .01 | .00 |
| 3 | .00 | .00 | | .00 | .00 | .00 | .00 | .00 | .00 | .00 | .00 | .00 | .00 | .00 | .00 |
| 4 | .00 | .00 | .00 | | .00 | .00 | .00 | .00 | .00 | .00 | .00 | .00 | .00 | .00 | .00 |
| 5 | .00 | .00 | .00 | .00 | | **.18** | .00 | .00 | .00 | .00 | .00 | .01 | .00 | .00 | .00 |
| 6 | .00 | .00 | .00 | .00 | .18 | | .00 | .00 | .00 | .02 | .04 | .00 | .00 | **.23** | .00 |
| 7 | .00 | .00 | .00 | .00 | .00 | .00 | | .00 | .00 | .00 | .00 | .00 | .00 | .00 | .00 |
| 8 | .00 | .00 | .00 | .00 | .00 | .00 | .00 | | .00 | .00 | .00 | .00 | .00 | .00 | .00 |
| 9 | .00 | .00 | .00 | .00 | .00 | .00 | .00 | .00 | | .00 | .00 | .00 | **.98** | .00 | .00 |
| 10 | .00 | .42 | .00 | .00 | .00 | .02 | .00 | .00 | .00 | | **.57** | .00 | .00 | **.33** | .00 |
| 11 | .00 | .27 | .00 | .00 | .00 | .05 | .00 | .00 | .00 | .57 | | .00 | .00 | **.16** | .00 |
| 12 | .00 | .00 | .00 | .00 | .01 | .00 | .00 | .00 | .00 | .00 | .00 | | .00 | .00 | **.90** |
| 13 | .00 | .00 | .00 | .00 | .00 | .00 | .00 | .00 | .98 | .00 | .00 | .00 | | .00 | .00 |
| 14 | .00 | .01 | .00 | .00 | .00 | .23 | .00 | .00 | .00 | .33 | .16 | .00 | .00 | | .00 |
| 15 | .00 | .00 | .00 | .00 | .00 | .00 | .00 | .00 | .00 | .00 | .00 | .90 | .00 | .00 | |

However, this is only a single series of experiments on 15 samples of tremorogramms. If we increase the number of series, which was made, there has been a marked statistical regularity, which presented in Table 5. From this table it follows that for a test subject without physical training $< k_{12} \approx 2 < k_{11}$.

**DISCUSSION**

The result of comparison of 15 series of TMG samples from two different groups of test subjects at the resting state of rest and 15 series with physical load (3*N*) shows that there is no statistical stability of samples of tremorogramms for woman or man. The repetition occurs without statistical "repetition", and tremorogramm samples are nearly all different, and it is impossible to obtain two consecutive identical samples (randomly!). We have a chaotic kaleidoscope of distribution functions *f(x)* for tremorogramms of every (one) test subject (in equilibriums state of all regulatory systems, like NMS). The average number of hits of random pairs of tremorogramm samples of test subjects (woman) $<k_{11}>=2.93$, which is significantly increase under physical load to $<k_{12}>=5.67$.

A different situation was observed for tested men, where $<k_{21}>=4.53$ less than $<k_{22}>=8.27$, but these differences more significant, that are the differences for women. The same tendency was observed in all subjects in the mode 225 repetitions of the measurement tremorogramma with a load (3N) and without a load, however, the values of $k_1$ and $k_2$ had individual character (in some subjects, $<k_1>=4$ and $<k_2>=7.4$ etc.). Our examples are typical for man and woman but we state that individual medicine (and psychology) needs of individual calculation quasi-attractors and matrix (like 1-4).

The results prove significant individual differences in the parameters of the tremor and questioned the appropriateness of combining different people in the statistical group in general . We now turn to a personalized medicine, where each person has their phase portrait in a limited (in size) of the phase space of states is just a vector of homeostasis $x=x(t)=(x_1,x_2,...,x_m)^T$, where m can be very large: $m>10$ or $m>100$, etc. For tremor including the phase coordinates are: $x_1(t)$ – coordinate, $x_2(t)=dx_1/dt$ is the velocity, $x_3(t)=dx_2/dt$ is the acceleration of the limb in space.

It was also revealed that the average number of hits $<k>$ for women and men in conditions of a rest differ somewhat (Table 5), which is a marker of fitness of man or woman of Russian North. The number of matches $<k_{21}>$ (man) is initially greater than $<k_{11}>$ (woman): $k_{11}=2.93 < k_{21}=4.53$. Accordingly increases the average number of hits $<k>$ in terms of static load (3*N*): for woman $k_{12}=5.67 < k_{22}=8.27$ (for man). Thus, the number of matches of arbitrary pairs of samples (*k*) for women and for men may be market of gender difference. This pattern was observed in all test subjects (12 women, 12 men).

Overall, this is an appropriate time to talk about the chaotic dynamics of tremor, although postural tremor occurs with the participation of consciousness (central nervous system) and it can be considered as a voluntary movement, but the reaction occurs randomly (without statistical stability) and then there are two fundamental problems of biomechanics and physiology of movements in general: 1) what is considered as voluntary motion (?); 2) what is the difference between voluntary and involuntary movements and how they can be measured (?). In authors' opinion, perhaps the way to solve these problems will be useful information-thermodynamic approach (see for example [ 38,39]).

*Table 5*

**The number of matches ($k_1$ and $k_2$) matrices for pairwise comparison of tremorogramm of test subjects in 15 series of experiments (Wilcoxon test, $p$<0.05)**

| № | Test subject FDU (woman) | | Test subject GDV (man) | |
|---|---|---|---|---|
| | Without load | With load 3*N* | Without load | With load 3*N* |
| 1 | 2 | 5 | 3 | 6 |
| 2 | 2 | 8 | 5 | 6 |
| 3 | 1 | 8 | 5 | 11 |
| 4 | 2 | 6 | 8 | 5 |
| 5 | 1 | 7 | 5 | 11 |
| 6 | 4 | 6 | 4 | 9 |
| 7 | 4 | 4 | 5 | 7 |
| 8 | 4 | 4 | 4 | 8 |
| 9 | 3 | 2 | 3 | 9 |
| 10 | 9 | 5 | 7 | 10 |
| 11 | 3 | 10 | 5 | 7 |
| 12 | 1 | 5 | 4 | 9 |
| 13 | 2 | 5 | 2 | 8 |
| 14 | 5 | 4 | 3 | 11 |
| 15 | 1 | 6 | 5 | 7 |
| **<k>** | **2.93** | **5.67** | **4.53** | **8.27** |

We came in this direction to another fundamental problem of physiology and Biomedicine: how quantitatively describe homeostasis (we are talking about not only homeostasis of NMS but about cardio-vascular systems (CVS) too [10-16, 26-35]) and what is homeostasis in general, if we don't have stationary states of NMS in the form $dx/dt=0$ ($x(t)$ varies continuously) and no statistical stability of distribution functions $f(x)$ for parameters of NMS? The solution to this problem is based on the test evaluations of the chaos of the parameter vector $x(t)$ [11, 13, 15, 23]. Recall that we have calculated the average number of hits of random pairs of samples ($<k>$) and standard deviation $\sigma$,± for all the 15 matrices with no load and under the impact of physical load (3N) of each of 30 test subjects (a man and a woman) [33-37].

This was manifested when it is almost impossible to get two consecutively registered tremorogramms where we would observe coincidences of $f(x)$, i.e., as a rule $f_j(x_i)\neq f_{j+1}(x_i)$ for any number of samples $j$. In case of tremor of any person (man or woman) the probability $p$ of coincidence of these features (i.e. $f_j(x_i)=f_{j+1}(x_i)$) does not exceed probability $p\leq 0.001$. This is an extremely small value and it proves existence of Eskov-Zinchenko effect [1-8. 17-24] and restricts the possibility of statistical description of the movements (in biomechanics). It requires a different mathematical apparatus and other methods to describe constant movements (or their changes).

In the framework of TCS, we constructed (for every testing subject) phase planes for all 225 samples, - all 15 series ($N$) of experiments for 15 samples in each ($n$) for each subject before and after a local cold exposure. The areas S of QA were calculated, which were found as the product of two variational ranges of the phase coordinates $\Delta x_1$ and $\Delta x_2$, that is $S=\Delta x_1 * \Delta x_2$. A typical example of phase trajectories without load and after the local cold impact is shown in figure 2.

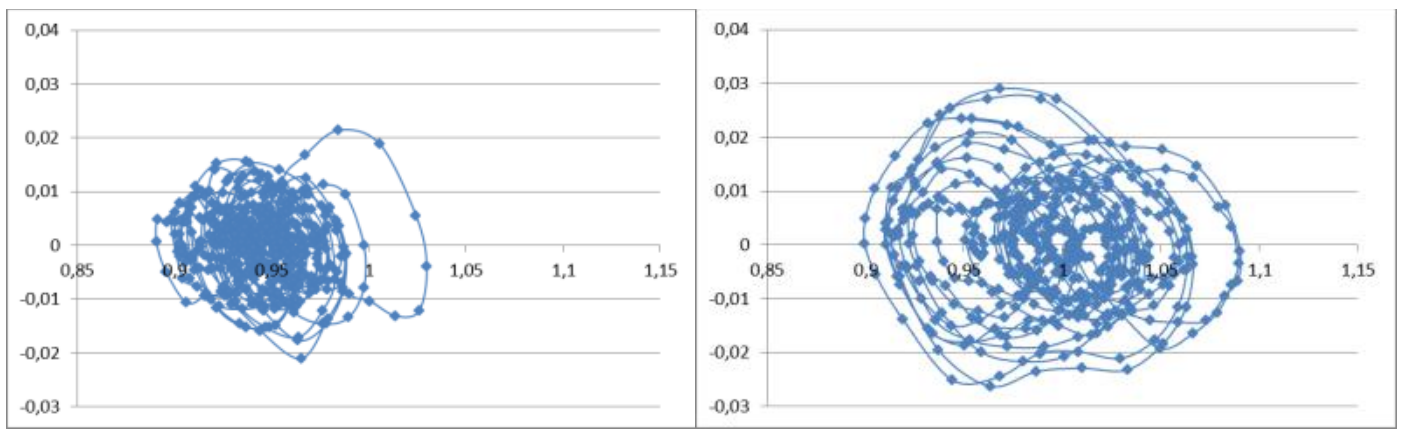

A Б

Figure 2. An example of the phase trajectories of one of the experiments of the subject GDV A – without ($F_1$=0) load ($S_{before}$=2.22* $10^{-6}$ c.u.); B – under local load ($F_2$=3$N$ ($S_{load}$=5.17*$10^{-6}$ c.u.).

In this case, the vector $x(t)=(x_1,x_2)^T$ performs chaotic motions within these QA (their S). An analysis of all the obtained values of S presents a similar picture in the form of data for each subject: separately for non-athletes and athletes. We emphasize that the statistical processing of the whole group does not make sense (the postulate of TCS), and we make comparisons for each subject separately for his area of QA (table 6), because we have a chaotic kaleidoscope of statistical distribution functions $f(x_i)$.

*Table 6*

**Areas ($S$*$10^{-6}$) of phase planes of samples of tremorograms of subjects**

| | Without load ($F_1$=0) | | With load ($F_2$=3$N$) | |
|---|---|---|---|---|
| № experiment | $S_1$, woman | $S_2$, man | $S_3$, woman | $S_4$, man |
| 1 | 0.35 | 0.45 | 0.33 | 0.98 |
| 2 | 0.13 | 0.19 | 0.39 | 0.11 |
| 3 | 0.14 | 0.29 | 0.29 | 0.68 |
| 4 | 0.26 | 0.10 | 0.31 | 0.45 |
| 5 | 0.39 | 0.26 | 0.53 | 0.53 |
| 6 | 0.11 | 0.44 | 0.83 | 0.67 |
| 7 | 0.82 | 0.20 | 0.39 | 0.97 |
| 8 | 0.82 | 0.22 | 0.17 | 0.72 |
| 9 | 0.29 | 0.29 | 0.31 | 0.97 |
| 10 | 0.38 | 0.16 | 0.46 | 0.46 |
| 11 | 0.18 | 0.13 | 0.80 | 0.41 |
| 12 | 0.13 | 0.67 | 0.11 | 0.94 |
| 13 | 0.23 | 0.19 | 0.32 | 0.48 |
| 14 | 0.18 | 0.17 | 0.89 | 0.29 |
| 15 | 0.19 | 0.11 | 0.29 | 0.58 |
| $<S>$ | 0.24 | 0.22 | 0.49 | 0.68 |

The Table 6 shows an example of the areas values of QA *S* for the samples of the tremorogramms of the subject FDU (women) and the GDV (men) for one of the experiment series: without load ($F_1$=0) and with load ($F_2$=3*N*). We propose to make repetitions of measurements and calculate the matrices of pair comparisons of samples before and after exposure (we had local cooling) to exit the emergent process in biomechanics and psychology. However, a simpler way is to calculate the parameters of phase planes of TMG. In all the studies that we carried out, the woman always showed a smaller value of the areas of QA OF TMG, than the man. In our examples, the average area $<S_2>$ for QA in the TMG for man is almost 2 times more than the average area $<S_1>$ for the woman ($<S_1>$ = 0.24, $<S_3>$ = 0.49), the medians give somewhat smaller effect, but it is also significant (0,48 and 0,31 respectively).

A similar (and steady) picture of differences is given by the area S for QA in TMG subjects under load $F_2$=3*N*. Here, $<S_2>$ = 0.22 for woman and $<S_4>$ = 0.68 for man, the size of the QA area quantitatively represents the NMS response to the stress-effect (under load $F_2$=3*N*). Calculation of QA parameters is a very effective method in biomechanics for evaluating the features of the regulation of motor functions (in our example these were involuntary movements in the form of postural tremor). Note that this approach gives a real quantitative assessment of the state of NMS, the work of the whole system of organization of movements, about which 70 years ago our outstanding predecessor N.A. Bernstein [25] tried to express as a hypothesis «about repetition without repetitions».

**CONCLUSION**

Thus, the calculation of matrices of pair comparisons of samples and parameters of QA for TMG makes it possible to evaluate not only the stationary state of homeostasis (in our case, for example, NMS), but also to distinguish these stationary states of homeostasis between individual subjects. In our case, we are talking about the homeostasis of men and women (with load and without load). We now demonstrate that the number of pairs of *k* coincidences of TMG samples in matrices (Tables 1-5) is different for the man and the woman.

Simultaneously, we observe pronounced changes in the numbers $k_2$ and $k_4$ in different subjects after a cold exposure to NMS. In this unidirectional changes in the area of QA, that in a calm state of athletes always have lower values than those without special physical training. Calculation of matrices and QA gives a new and effective method for diagnosing the condition of subjects who are in different physiological and mental states. Against the background of statistical chaos of TMG samples, we can observe certain patterns within the framework of the theory of chaos - self-organization.

Therefore, all samples are different (at least 10% of TMG pairs can not show statistical coincidence). It is completely incomprehensible what to take for a change under the influence of load, if without load all successively received tremorograms from one subject (being in one homeostasis) are significantly different. The hypothesis of N. A. Bernstein was realized in the effect of Eskov-Zinchenko (there are no coincidences in a row of received samples). The quasiattractors value demonstrate the homeostatic state for equilibrium state of NMS and the difference between areas $S_1$, $S_2$, $S_3$, $S_4$ demonstrate the difference of homeostasis and the gender specificity of NMS.

**Gratitude**

The work is executed at support of RFFI grant № A 18-07-00162, A 18-07-00161